\documentclass{article}

\usepackage{booktabs}
\usepackage{multirow}
\usepackage{PRIMEarxiv}
\usepackage{multirow}
\usepackage[table,xcdraw]{xcolor}
\usepackage[utf8]{inputenc} 
\usepackage[T1]{fontenc}    
\usepackage{hyperref}       
\usepackage{url}            
\usepackage{booktabs}       
\usepackage{amsfonts}       
\usepackage{nicefrac}       
\usepackage{microtype}      
\usepackage{lipsum}
\usepackage{fancyhdr}       
\usepackage{graphicx}       
\graphicspath{{media/}}     

\title{Adversarial Masked Image Inpainting for Robust Detection of Mpox and Non-Mpox
}

\author{
  Yubiao Yue \\
  Guangzhou Medical University \\
  \texttt{jiche2020@126.com} \\
   \And
Zhenzhang Li \\
Guangdong Polytechnic Normal University \\
  \texttt{zhenzhangli@gpnu.edu.cn} \\
}

\begin{document}
\maketitle

\begin{abstract}
Due to the lack of efficient mpox diagnostic technology, mpox cases continue to increase. Recently, the great potential of deep learning models in detecting mpox and non-mpox has been proven. However, existing models learn image representations via image classification, which results in they may be easily susceptible to interference from real-world noise, require diverse non-mpox images, and fail to detect abnormal input. These drawbacks make classification models inapplicable in real-world settings. To address these challenges, we propose "Mask, Inpainting, and Measure" (MIM). In MIM's pipeline, a generative adversarial network only learns mpox image representations by inpainting the masked mpox images. Then, MIM determines whether the input belongs to mpox by measuring the similarity between the inpainted image and the original image. The underlying intuition is that since MIM solely models mpox images, it struggles to accurately inpaint non-mpox images in real-world settings. Without utilizing any non-mpox images, MIM cleverly detects mpox and non-mpox and can handle abnormal inputs. We used the recognized mpox dataset (MSLD) and images of eighteen non-mpox skin diseases to verify the effectiveness and robustness of MIM. Experimental results show that the average AUROC of MIM achieves 0.8237. In addition, we demonstrated the drawbacks of classification models and buttressed the potential of MIM through clinical validation. Finally, we developed an online smartphone app to provide free testing to the public in affected areas. This work first employs generative models to improve mpox detection and provides new insights into binary decision-making tasks in medical images.
\end{abstract}


\section{Introduction}
Mpox (Monkeypox), is a zoonotic viral disease caused by the mpox virus \cite{1}. This disease is transmitted through respiratory droplets, direct contact, or contaminated objects \cite{2} and results in symptoms such as fever, rash, and swollen lymph nodes in patients \cite{3}. Among various symptoms, the rash is a characteristic feature of mpox patients \cite{4} and visually resembles chickenpox and measles \cite{5}. Statistics show that the fatality rate of mpox varies between 0\% and 11\% in the general population, with a greater rate among small children \cite{6}. Initially, mpox was primarily endemic to remote regions of west and central African countries \cite{7}. However, since May 2022, mpox has continuously swept through many countries and was urgently listed as a "Public Health Emergency of International Concern" by the World Health Organization. As of September 19, 2023, 115 countries have reported 90,439 cases of mpox \cite{8}.

Like other infectious diseases, timely and rapid detection of mpox patients is key to control outbreaks \cite{9}. Especially in this fast-paced era with frequent interpersonal interactions, the spread of infectious diseases can quickly become global \cite{10}. Therefore, routine monitoring and management of the mpox epidemic are indispensable, underscoring the real need for more efficient detection techniques \cite{11}. Utilizing Polymerase Chain Reaction (PCR) to detect the virus's DNA from a patient's rash is the preferred laboratory method for mpox diagnosis. However, the PCR requires advanced specialized equipment and experienced personnel and can be costly. Such costs are often prohibitive for resource-limited areas with higher mpox incidence rates \cite{12}. Moreover, existing detection techniques are ill-suited for mass screening during outbreaks. The abovementioned dilemmas raise a question: Is there a more efficient and cost-effective detection strategy?

The rapidly advancing deep learning technology points the way forward. Since the outbreak of mpox last year, an increasing number of researchers have turned their attention to this disease. They have successfully demonstrated that various deep-learning models in computer vision can efficiently diagnose mpox using rash images. However, our review suggests that these studies have yet to profoundly consider models' detection capabilities in real-world settings. Specifically, the first step in the typical workflow of developing these models is to collect images of mpox. Researchers then aggregate various non-mpox images and train a binary-class classification model \cite{13,14,15,16,17,18,19,20,21,22,23,24} to distinguish between mpox and non-mpox. Alternatively, researchers subdivide non-mpox data and train multi-class classification models to differentiate between mpox, chickenpox, measles, normal skin, and other skin diseases \cite{25,26,27,28,29,30}. While these models have shown preliminary efficacy, they have the following drawbacks. Firstly, during classification tasks, to quickly optimize parameters, models tend to find a "shortcut" \cite{31}, which might lead to inadequate learning of critical features in the training set and susceptibility to noise \cite{32,33}. Secondly, the detection performance of the classification model relies not only on numerous mpox images but also on a diverse set of non-mpox images. However, procuring comprehensive medical images remains a formidable challenge. Besides, estimating the number and categories that non-mpox images should contain is difficult. Lastly and most crucially, classification models do not refuse to detect when confronted with an input not belonging to known categories. In other words, classification models always expect that the input image belongs to the training set categories. However, in the real world, the number of patients with non-mpox rashes vastly outnumbers those with mpox, and their rashes often do not belong to the model's training set category. Classification models cannot refuse to detect these patients and will unthinkingly judge the model input as one of the classes in the training set, which greatly reduces the safety and usefulness of these models.

To solve these challenges, our work aims to develop a deep learning model capable of robustly detecting mpox and non-mpox in real-world environments. We summarize our contributions as follows: (1) We analyzed the drawbacks of classification models used to detect mpox and non-mpox. We have defined this detection task as a One-class Novelty Detection Task in computer vision. In this scenario, mpox data is treated as normal, and all non-mpox data from the real world is treated as abnormal. (2) We proposed a novel detection method called "Mask, Inpainting, and Measure" (MIM). The core of MIM is that a Generative Adversarial Network (GAN) models normal images via image inpainting and a masking strategy. Since the model is trained only on mpox images, it cannot effectively repair anomalous images during inference. Thus, new images can be classified as mpox or non-mpox based on the similarity between the restored and original images. This method cleverly detects mpox versus non-mpox without needing any non-mpox datasets. Importantly, MIM is not easily disturbed by various anomalous data in the real-world environment. (3) To our knowledge, this is the first time that generative models have been used for detecting mpox versus non-mpox. To enhance the detection performance of MIM, we compared various image similarity calculation methods. We found that using both the structural similarity index measure and the histogram Bhattacharyya distance resulted in the best detection performance of MIM. (4) We demonstrated the robustness and potential of MIM by comparing it with some popular classification models. Besides, we also revealed the limitations of classification models. (5) We further validated MIM using non-mpox images collected in clinical settings and developed a smartphone application to provide a more convenient and efficient testing service for the public in areas affected by mpox and resource-limited regions.

\section{Materials and Methods}
\subsection{Novelty Detection}
Novelty detection (ND) is a classical and important research problem in machine learning with a wide range of applications, including medical diagnosis, fraud detection, and autonomous driving \cite{34}. It aims to detect any test samples that do not belong to any training category \cite{35}. In this task, a sample similar to the training data can be termed a "Normal sample" or "In-distribution sample." Otherwise, it is called an "Anomaly sample" or "Out-of-distribution sample" \cite{34,35}. Based on the number of training classes, ND contains two different settings. The first is one-class novelty detection (one-class ND), which means only one class exists in the training set. The second is multi-class novelty detection (multi-class ND), which means multiple classes exist in the training set \cite{36}.

Our research falls under one-class novelty detection. Traditionally, One-class ND is viewed as a representation-learning problem \cite{37}. In this task, the model solely attempts to capture the distribution of normal samples, eventually detecting unknown anomalies or new concepts. Earlier, related algorithms largely depended on Support Vector Machines (SVM). With the advent of deep learning techniques in recent years, researchers have shifted focus toward learning a representative latent space of in-distribution data \cite{38}. As a potent unsupervised representation learning method, autoencoder has demonstrated immense potential in modeling images \cite{39}. Following this, autoencoder-based approaches have taken a dominant position in the field, such as Denoising Autoencoders (DAE) \cite{40}, Variational Autoencoders (VAE) \cite{41}, and Adversarial Autoencoders (AAE) \cite{42}. Specifically, to detect out-of-distribution images, an autoencoder is first trained on in-distribution samples. At test time, the reconstruction error of a sample serves as an anomaly score. Ultimately, the classification of an image is determined based on this anomaly score. This idea stems from an intuition: since the model only learns the distribution of normal samples, the reconstruction error for normal samples will be lower than that of anomalous samples.
\subsection{Data Source}
Based on the requirements of the novelty detection task, we need to collect mpox images as in-distribution data for model training and gather various types of non-mpox images as anomaly data to verify the effectiveness of the proposed method. The data used in this work comes from four sources. The first part is MSLD \cite{43,44}. It is a recognized and expert-verified mpox dataset. It contains images of mpox, chickenpox, hand-foot-mouth disease (HFMD), measles, and normal skin. In practical settings, patients or doctors should take close-up front-facing photographs of skin lesions and submit them to the visual skin disease detection system. Therefore, we exclude samples that do not meet the actual shooting angle and distance, have complicated backgrounds, are repetitive, and are blurred. Figure 1a illustrates our requirements for image quality. Figure 1b reports the number of images in different categories.

\begin{figure}
    \centering
    \includegraphics[width=\textwidth]{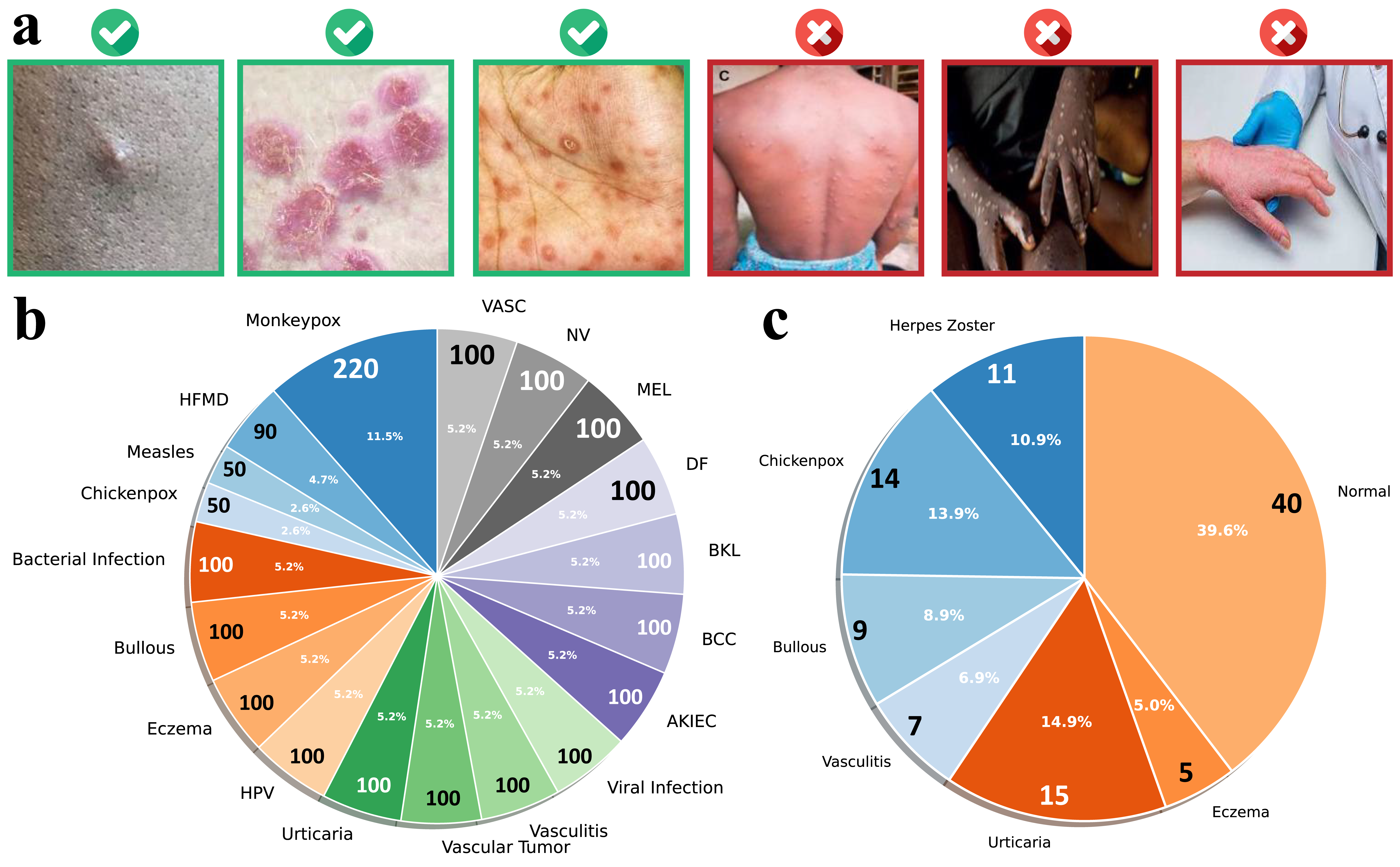}
    \caption{Requirements for image quality and the distribution of data used in this work. a: Images with green borders represent qualified images, while images with red borders represent unqualified images; b: The number of images in each category included in the data from the first part to the third part; c: The number of images in each category included in the fourth part.}
    \label{fig1}
\end{figure}
The second part is the ISIC 2018 dataset \cite{45,46}. This dataset contains seven types of skin diseases: Melanoma, Melanocytic nevus, Basal cell carcinoma, Actinic keratosis / Bowen’s disease (intraepithelial carcinoma), Benign keratosis (solar lentigo / seborrheic keratosis / lichen planus-like keratosis), Dermatofibroma, and Vascular lesion. The corresponding abbreviations for these seven diseases are MEL, NV, BCC, AKIEC, BKL, DF, and VASC. In this study, we randomly selected 100 images from each category. The third part is the Dermnet dataset \cite{47} from the Kaggle website, which includes 23 types of skin diseases. We selected eight skin diseases (Bacterial Infection, Bullous, Eczema, HPV, Urticaria, Vascular Tumor, Vasculitis and Viral Infection) from it that have features similar to mpox. We filtered out 100 images from each category based on the image quality. Figure 1b reports the number of images for each type of skin disease from the first to the third part of the data. Figure 2 displays samples of skin diseases from the first part to the third part.
\begin{figure}
    \centering
    \includegraphics[width=\textwidth]{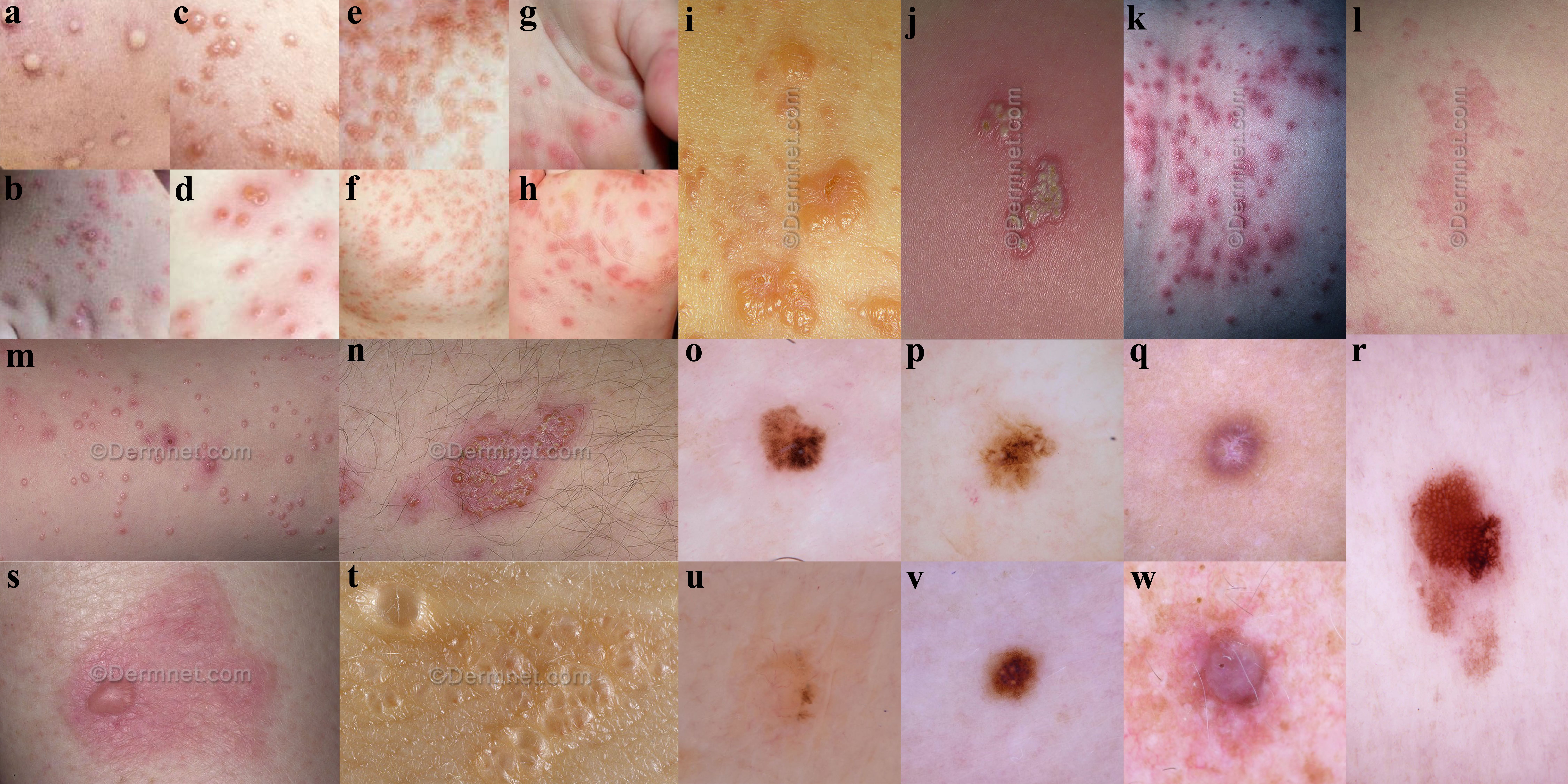}
    \caption{Samples of skin diseases included in first part to third part data. a-b: Mpox; c-d: Chickenpox; e-f: Measles; g-h: Hand, foot and mouth disease; i: Bullous; j: Viral infection; k: Bacterial infection; l: Urticaria; m: HPV; n: Eczema; o: AKIEC; p: BKL; q: DF; r: MEL; s: Vasculitis; t: Vascular tumor; u: BCC; v: NV; w: VASC.}
    \label{fig2}
\end{figure}

The fourth part is the clinical dataset. The Department of Dermatology at the Second Affiliated Hospital of Guangzhou Medical University provided us with skin lesion images of 60 patients. According to local legislation and institutional requirements, the clinical data used in this study did not require approval from the ethics committee. The Helsinki Declaration's tenets were scrupulously followed throughout the study to respect the rights, privacy, and anonymity of the subjects. The use of de-identified data precluded the use of informed consent. 

Additionally, we collected 40 high-definition images of normal skin from team members using smartphones based on a reasonable shooting angle. Figure 1c displays the number of images corresponding to each type of skin disease in the clinical data. Figure 3 shows some rash samples provided by the hospital and normal skin samples collected by ourselves.

\begin{figure}
    \centering
    \includegraphics[width=\textwidth]{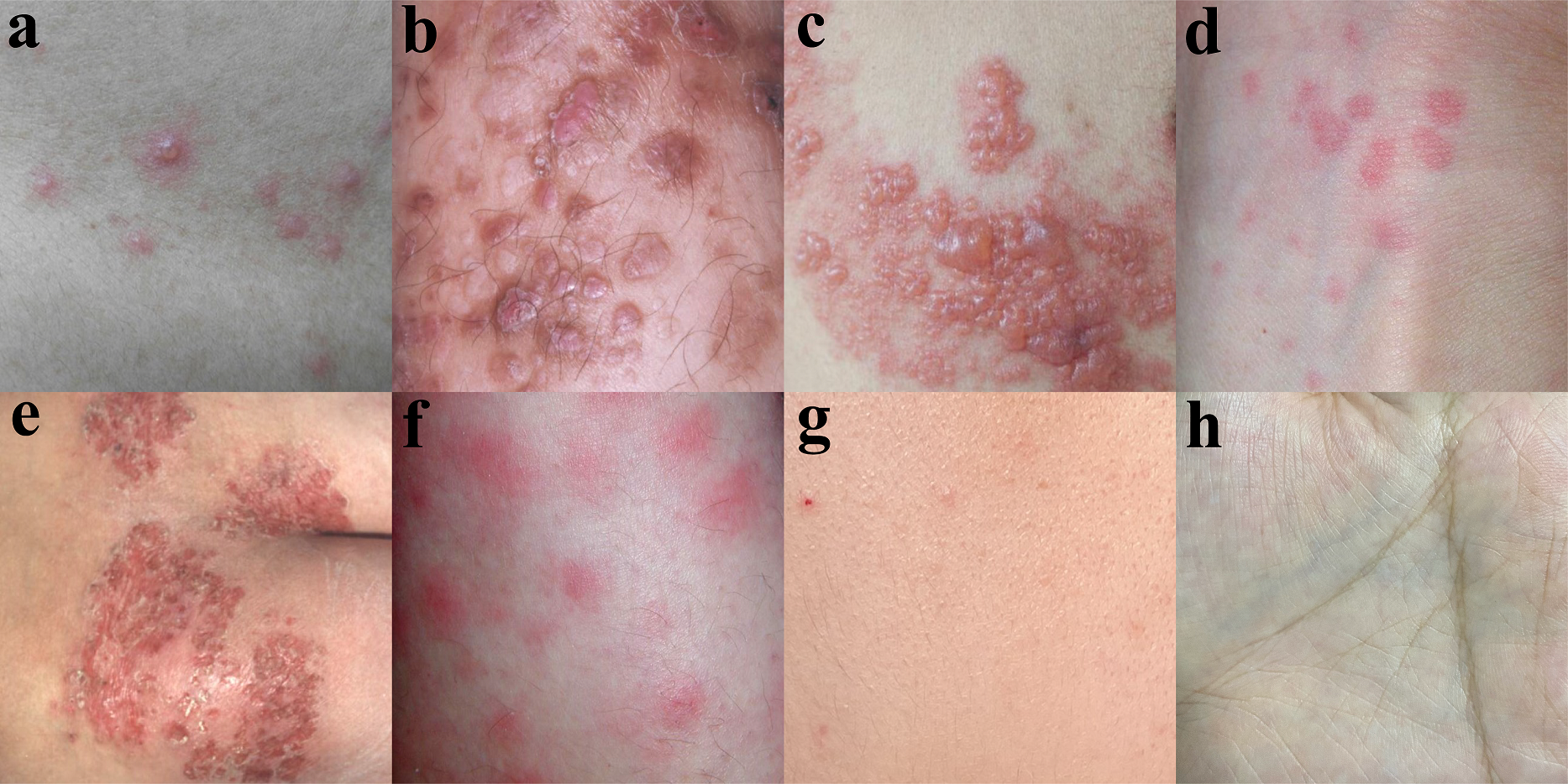}
    \caption{Skin disease samples provide by the hospital and normal skin samples collected by ourselves. a: Chickenpox; b: Bullous; c: Herpes zoster; d: Urticaria; e: Eczema; f: Vasculitis; g-h: Normal skin.}
    \label{fig3}
\end{figure}

\subsection{Mask, Inpainting and Measure}
This paper proposes "Mask, Inpainting, and Measure" (MIM) to detect monkeypox and non-monkeypox. The core idea of MIM is from the autoencoder-based reconstruction method. Specifically, MIM first uses a mask to destroy the image's spatial structure and semantic information, and then uses a generative model, namely the Generative Adversarial Network (GAN), to inpaint the image. Finally, MIM employs an image similarity metric to measure the similarity between the original and repaired images. This idea stems from the following intuition: Since the GAN only repairs masked monkeypox images during training, it only learns the contextual and semantic information of the monkeypox images. When a non-monkeypox image is input into the network, the network cannot repair it well, resulting in an image dissimilar to the original image. In the end, the similarity between the original image and the repaired image is used as an anomaly score, and the category of the image is determined through a threshold. In the following sections, we will individually introduce the core details of MIM.
\subsubsection{Mask}
In computer vision, a "mask" typically refers to a binary image where certain regions are marked as 1 and others as 0. Recent literature has demonstrated that masked image modeling can significantly improve a model's capability to learn data representations \cite{48,49}. The principle behind this strategy is to obscure a portion of the input image and have the deep network predict the masked signal based on visible cues. This idea originated from masked signal modeling, which can force the network to predict the obscured parts, thereby aiding the model in learning a more robust and richer data representation.

In this work, we utilize a rectangular white mask. Without considering the image channels, we define the following notation. $I$ denotes an image of size $H \times W$ ($H$ and $W$ represent the image's height and width, respectively). $I(h,w)$ denotes the pixel value at each position of $I$. After normalizing $I$, $I(h,w)$ lies in the range $[0,1]$. Next, we use $M$ to denote a mask of size $H \times W$ and choose the coordinate range of the masked area, i.e., $h_1 \sim h_2 (0 \leq h_1 \leq h_2 \leq H)$ and $w_1 \sim w_2 (0 \leq w_1 \leq w_2 \leq W)$, and the pixel values of each position in $M$ can be expressed using the following formula:
\begin{equation}
\label{eq1}
M(h,w) = \left\{ \begin{array}{l}
1,{h_1} \le h \le {h_2}{\rm{ }},{\rm{ }}{w_1} \le w \le {w_2}\\\\
0,otherwise
\end{array} \right.
\end{equation}
Last, we use I' to denote the masked image and the following formula is established:
\begin{equation}
\label{eq2}
I'(h,w) = I(h,w) \odot (1 - M(h,w)) + M(h,w)
\end{equation}
, where $ \odot $ denotes the Hadamard product. In terms of how masks are used, MIM uses a random mask strategy in the training pipeline (Figure 4a), that is, the position of the mask is not fixed, which can improve the robustness of the generator. In the inference pipeline, MIM adopts a center masking strategy (Figure 4b)

\subsubsection{Inpainting}
\begin{figure}
    \centering
    \includegraphics[width=\textwidth]{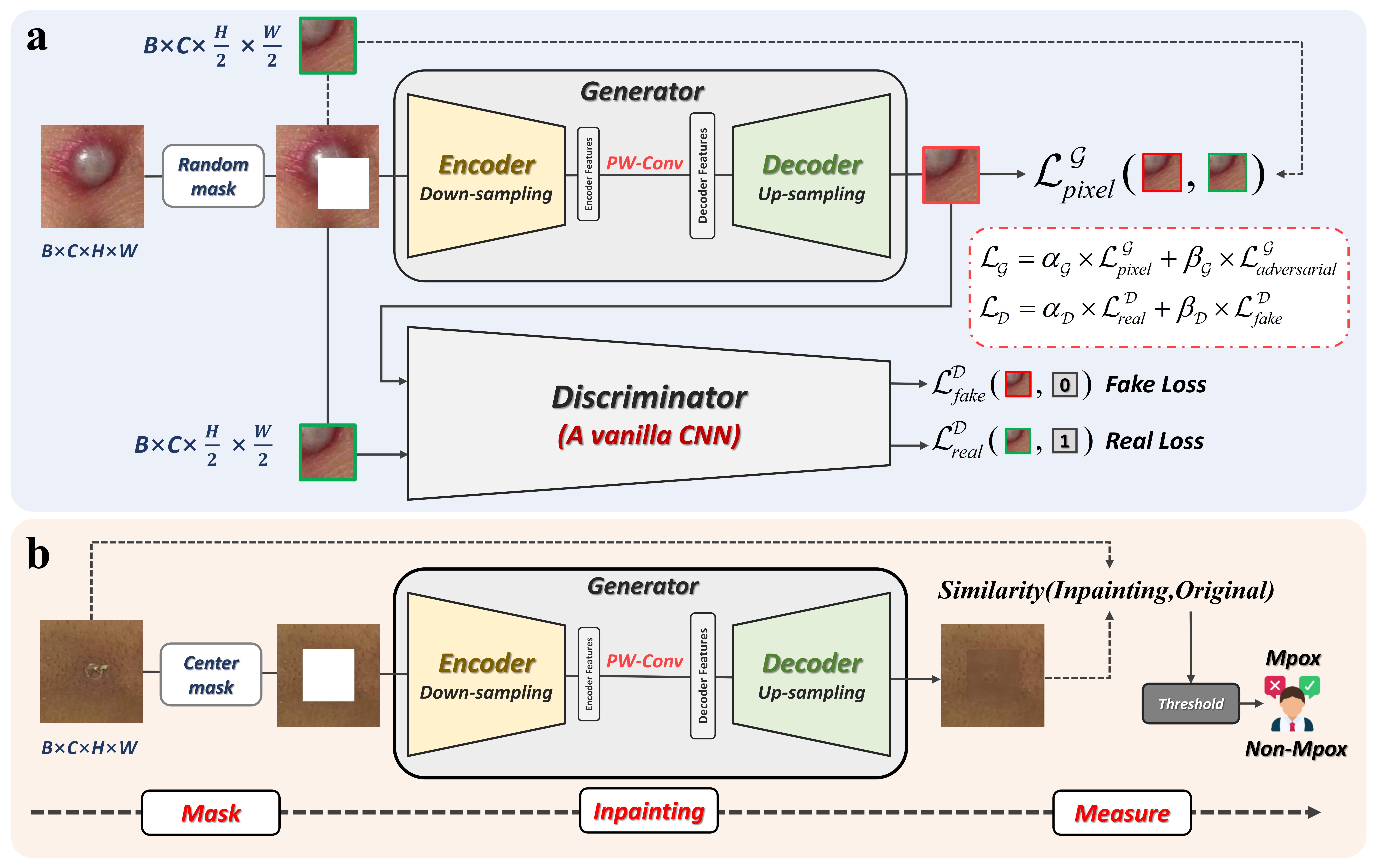}
    \caption{The workflow of MIM. Here, PW-Conv means Point-wise Convolution and CNN means a Convolutional Neural Network. a: MIM's training pipeline; b: MIM's inference pipeline.}
    \label{fig4}
\end{figure}
In the MIM's training pipeline (Figure 4a), a generative model learns the key representations of training data (in-distribution data) by predicting the masked region of an input image. In this work, we adopted the generative adversarial network as the generative model of MIM \cite{50}. Specifically, we utilized a vanilla autoencoder as a generator to inpaint the masked region in the original image. We employed a vanilla convolutional neural network as a discriminator to distinguish the masked regions predicted by the autoencoder. The output of the discriminator will be a part of the overall loss of the generator, thereby effectively improving the generator's robustness and generalization. Figure 4 illustrates the detailed process of the MIM. It is worth noting that any generative network architecture can replace the Generator in MIM. Similarly, the architecture of the discriminator is not restricted. Table S1 and Table S2 show the internal structure of the generator and discriminator, respectively.
We used a weighted loss function to optimize the model parameters of the autoencoder. Specifically, the loss function of the generator comes from two parts: pixel-level loss and adversarial loss. We formalize the calculation process of loss function: ${\rm{{\cal G}}}(x;{\theta _{\rm{{\cal G}}}})$ and ${\rm{{\cal D}}}(x;{\theta _{\rm{{\cal D}}}})$ denotes generator and discriminator, respectively. Given a distribution of interest ${D_{in}}$, we use ${I\sim{D_{in}}}$ to denote the MIM's input. After using a mask, ${I_{masked}}$ denotes the masked area in $I$ and $I'$ denotes the masked image, respectively. Then, the output of the ${\rm{{\cal G}}}(x;{\theta _{\rm{{\cal G}}}})$ the masked area that needs to be predicted, is denoted by ${\rm{{\cal G}}}(I';{\theta _{\rm{{\cal G}}}})$. Last, the pixel-level loss of the generator can be defined as follows:
\begin{equation}
\mathrm{{\mathcal{L}}}_{\mathrm\textbf{{pixel}}}^{\mathrm{\mathcal{G}}} = \mathbb{E}_{I\sim D_{\mathrm{in}}}\left[ f_{MAE}(\mathrm{\mathcal{G}}(I';\theta_{\mathrm{\mathcal{G}}}),I_{\mathrm\textbf{{masked}}}) \right]
\end{equation}
$f_{MAE}$ in the above formula represents the mean absolute value error between real masked area and fake masked area, and its formula is as follows:
\begin{equation}
{f_{MAE}}({x_1},{x_2}) = \frac{1}{{C \times H \times W}}\sum\nolimits_{c = 1}^C {\sum\nolimits_{h = 1}^H {\sum\nolimits_{w = 1}^W {\left| {{x_1}(c,h,w) - {x_2}(c,h,w)} \right|} } }
\end{equation}
, where $x1$ and $x2$ respectively denotes images with the same shape. In addition, $C$, $H$, and $W$ represent the number of channels, height, and width of the image, respectively. $(c, h, w)$ represents the position of each pixel in the image.
The formula of the adversarial loss of generator is as follows:
\begin{equation}
\mathcal{L}_{adversarial}^{\mathcal{G}} = \mathbb{E}_{I\sim D_{\mathrm{in}}}\left[ f_{\mathrm{MSE}}\left(\mathcal{D}(\mathcal{G}(I',\theta_{\mathcal{G}});\theta_{\mathcal{D}}),\mathcal{T}_{0}\right) \right]
\end{equation}
Here ${\mathcal{T}}_0$ represents a matrix whose elements are all 0, and its shape and size are consistent with the output of the discriminator. $f_{MSE}$ represents mean square error, and its formula is as follows:
\begin{equation}
{f_{MSE}}({x_1},{x_2}) = \frac{1}{{C \times H \times W}}\sum\nolimits_{c = 1}^C {\sum\nolimits_{h = 1}^H {\sum\nolimits_{w = 1}^W {{{\left( {{x_1}(c,h,w) - {x_2}(c,h,w)} \right)}^2}} } }
\end{equation}
Finally, the total loss of the generator is calculated using a weighted sum:
\begin{equation}
\mathcal{L}_{\mathcal{G}} = \alpha_{\mathcal{G}} \times \mathcal{L}_{\mathrm{pixel}}^{\mathcal{G}} + \beta_{\mathcal{G}} \times \mathcal{L}_{\mathrm{adversarial}}^{\mathcal{G}}
\end{equation}
, where $\alpha_{\mathcal{G}}$ and $\beta_{\mathcal{G}}$ represent weight factors respectively. We set them to 0.01 and 0.99 respectively.

For the discriminator, we also used a weighted loss function. Specifically, the loss of the discriminator must consider the loss values from both real masked area and fake masked area predicted by the generator. The loss function of fake masked area is as follows:
\begin{equation}
\mathcal{L}_{\mathrm{fake}}^{\mathcal{D}} = \mathcal{L}_{\mathrm{adversarial}}^{\mathcal{G}}
\end{equation}
For real masked area, the loss function is:
\begin{equation}
\mathrm{{\mathcal{L}}}_{\mathrm\textbf{{real}}}^{\mathrm{\mathcal{D}}} = \mathbb{E}_{I\sim D_{\mathrm{in}}}\left[f_{MSE}(\mathrm{\mathcal{D}}(I';\theta_{\mathrm{\mathcal{D}}}),\mathcal{T}_{1}\right)]
\end{equation}
Here ${\mathcal{T}}_1$ represents a matrix whose elements are all 1, and its shape and size are consistent with the output of the discriminator. Based on the above, the total loss of the discriminator is:
\begin{equation}
\mathcal{L}_{\mathcal{D}} = \alpha_{\mathcal{D}} \times \mathcal{L}_{\mathrm{real}}^{\mathcal{D}} + \beta_{\mathcal{D}} \times \mathcal{L}_{\mathrm{fake}}^{\mathcal{D}}
\end{equation}
, where $\alpha_{\mathcal{D}}$ and $\beta_{\mathcal{D}}$ represent weight factors respectively. We set them to 0.5 and 0.5 respectively.
\subsubsection{Measure}
According to the novelty class detection process, we use the similarity between the original and inpainted images as an anomaly score. The similarity value determines whether the image belongs to mpox or non-mpox. In this work, we chose five metrics to calculate similarity.

Mean Absolute Error (MAE). Mean Absolute Error is a metric to measure the average of absolute errors between two sets of values. When comparing two images, it measures the average pixel absolute difference between the two images. The smaller the mean absolute error, the higher the similarity. Based on the formula for calculating the mean absolute error, the similarity can be calculated using the following formula:
\begin{equation}
Similarit{y_{MAE}} = 1 - {f_{MAE}}(x1,x2)    
\end{equation}
This formula shows that the smaller the mean absolute error, the higher the similarity between images.

Mean Squared Error (MSE). The Mean Squared Error is a metric that calculates the average of the squared differences between the corresponding pixels of two images. Like the Mean Absolute Error (MAE), it measures the difference between two sets of data, but by squaring the differences, it gives more weight to more significant errors, making it more sensitive to outliers. Based on the formula for calculating the mean squared error, the similarity can be calculated using the following formula:
\begin{equation}
    Similarit{y_{MSE}} = 1 - {f_{MSE}}(x1,x2)
\end{equation}
This formula shows that the smaller the mean square error, the higher the similarity between images.

Structural Similarity Index Measure (SSIM) 
\cite{51}. Structural Similarity Index Measure is a metric used to measure the similarity between two images. It provides a more comprehensive and accurate assessment of perceived quality differences between images compared to traditional metrics such as Mean Squared Error (MSE) or Mean Absolute Error (MAE). The main idea behind SSIM is to consider changes in structural information, luminance, and contrast, which are believed to be highly important in the human perception of image quality. When SSIM returns a value of 1, it indicates that the two images being compared are identical in terms of structural, luminance, and contrast characteristics. A value of 0 or near 0 indicates that the images have significant differences in these characteristics. The calculation formula of SSIM is as follows:
\begin{equation}
Similarit{y_{SSIM}}({x_1},{x_2}) = \frac{{(2{\mu _{{x_1}}}{\mu _{{x_2}}} + {c_1})(2{\sigma _{{x_1}{x_2}}} + {c_2})}}{{(\mu _{{x_1}}^2 + \mu _{{x_2}}^2 + {c_1})(\sigma _{{x_1}}^2 + \sigma _{{x_2}}^2 + {c_2})}}
\end{equation}
In this formula, $\mu_{x1}$ and $\mu_{x2}$ represent the mean of images $x1$ and $x2$, respectively. $\sigma_{x1}$ and $\sigma_{x2}$ represent the variance of images $x1$ and $x2$, respectively. $\sigma_{x1x2}$ represents the covariance of $x1$ and $x2$. $c1$ and $c2$ are two constants to avoid division by zero. $c1$ and $c2$ are defined as follows: ${c_1} = {({k_1}L)^2},{c_2} = {({k_2}L)^2}$, where K1 is 0.01, K2 is 0.03 and L is the dynamic range of pixel values.

Histogram Bhattacharyya Distance (HBD). A histogram is a tool used to represent the distribution of pixel intensities in an image. Through the histogram, we can understand attributes such as an image's brightness and color distribution. The Bhattacharyya distance is used to gauge the similarity between two probability distributions. When we apply the Bhattacharyya distance to the histograms of two images, it can measure the similarity between these two images in terms of color and brightness distribution, especially in terms of overall color distribution. Based on the principle of Bhattacharyya distance, we define the similarity as follows:
\begin{equation}
Similarit{y_{HBD}}({x_1},{x_2}) = 1 + ln(\sum\nolimits_i {\sqrt {{x_1}(i).{x_2}(i)} } )
\end{equation}
, where x1 and x2 respectively represent the histograms of the two images and $\sum$ is over all bins of the histograms.

In addition to the above four calculation methods, we intuitively propose a joint metric that simultaneously considers the HBD and SSIM between images. The formula of joint metric is as follows:
\begin{equation}
Similarit{y_{SSIM + HBD}}({x_1},{x_2}) = {\lambda _1} \times Similarit{y_{SSIM}}({x_1},{x_2}) + {\lambda _2} \times Similarit{y_{HBD}}({x_1},{x_2})
\end{equation}
In the formula, $\lambda1$ and $\lambda2$ represents weights, and we set them to 0.5 and 0.5 respectively.

To thoroughly and objectively evaluate the performance and potential of MIM, we followed the rules of previous work[34], [35], [37] and used the area under the receiver operating characteristic curve (AUROC) as a threshold-free evaluation metric.

\section{Results and Discussion}
\subsection{Implementation Details}
We conducted the entire experiment on a computer with an Nvidia GeForce RTX 4090 graphics card and Ubuntu 20.04 system. We used the Adam optimizer to optimize the parameters of each model in the MIM's training pipeline. In the Adam optimizer, $\beta$1 and $\beta$2 were set to 0.9 and 0.999, respectively. We set the initial learning rate to 0.001, the batch size to 128, and the number of epochs to 300. MIM's input size and mask size are 224×224×3 and 112×112×3, respectively. To prevent overfitting, we employed a learning rate decay strategy. Notably, we used five-fold cross-validation to avoid data randomness and thoroughly evaluate MIM. Specifically, there are 176 monkeypox images and 44 monkeypox images in each dataset. One thing to note is that we did not use any data augmentation strategies on the training set. However, when calculating the AUROC of MIM, we generated 220 new images by augmenting (horizontal flipping + vertical flipping + saturation change) the monkeypox images in each test set. The purpose of this is two-fold. First, monkeypox images in practical settings are relatively rare, so we tried to explore the representation-learning ability of MIM in a small-scale dataset. Second, there is a sizeable quantitative gap between the monkeypox images in the test set and the non-monkeypox images from out-of-distribution, which may cause AUROC not fully to reflect MIM's performance. Finally, we utilized 220 normal monkeypox images and 1690 abnormal non-monkeypox images to calculate AUROC. The average AUROC was used to evaluate MIM's final performance. Table 1 lists the detailed data usage.
\begin{table}[]
\centering
\caption{Distribution of data used to train and test MIM. ID and OOD mean In-distribution and out-of-distribution, respectively.}
\resizebox{\textwidth}{!}{%
\begin{tabular}{@{}lllll@{}}
\toprule
      & Train(ID)       & Test(ID)       & Real Test (ID)             & Reak Test (OOD)      \\ \midrule
Fold1 & 176 mpox images & 44 mpox images & (44) → (220) mpox   Images &                      \\
Fold2 & 176 mpox images & 44 mpox images & (44) → (220) mpox   Images &                      \\
Fold3 & 176 mpox images & 44 mpox images & (44) → (220) mpox   Images & 1690 non-mpox Images \\
Fold4 & 176 mpox images & 44 mpox images & (44) → (220) mpox   Images &                      \\
Fold5 & 176 mpox images & 44 mpox images & (44) → (220) mpox   Images &                      \\ \bottomrule
\end{tabular}
}
\end{table}
\subsection{The Best Similarity Metric}
We first explored the impact of various similarity metrics for MIM. Table 2 reports the AUROC calculated from different metrics. Among all the metrics we investigated, the metric we proposed exhibited the best performance, with an average AUROC of 0.8237(±0.0101) and up to 0.8401 in the best case. Additionally, the values for SSIM and HD were 0.7582(±0.0455) and 0.7979(±0.0243) respectively. The performance of pixel-difference-based metrics, MAE and MSE, was relatively poor, with values of 0.7214(±0.0312) and 0.6984(±0.0286), respectively.

\begin{table}[]
\centering
\caption{Differences in effects of different image similarity calculation methods.}
\resizebox{\textwidth}{!}{%
\begin{tabular}{@{}llllll@{}}
\toprule
           & SSIM+HBD        & SSIM            & HBD             & MAE             & MSE             \\ \midrule
Fold1      & 0.8123          & 0.6757          & 0.8395          & 0.6832          & 0.6685          \\
Fold2      & 0.8401          & 0.7660          & 0.8082          & 0.6935          & 0.6646          \\
Fold3      & 0.8167          & 0.7500          & 0.7876          & 0.7285          & 0.6994          \\
Fold4      & 0.8190          & 0.7989          & 0.7685          & 0.7302          & 0.7230          \\
Fold5      & 0.8303          & 0.8004          & 0.7855          & 0.7715          & 0.7364          \\
Mean(±Std) & 0.8237(±0.0101) & 0.7582(±0.0455) & 0.7979(±0.0243) & 0.7214(±0.0312) & 0.6984(±0.0286) \\ \bottomrule
\end{tabular}
}
\end{table}

We offer the following explanations regarding the differences in these results: When analyzing human rash images involving structure, texture, and color distribution characteristics, it is often not enough to measure the similarity between images only by the difference in pixel values. During the evaluation process, SSIM and HBD consider more of the images' visual and structural properties, making them more sensitive to the rich texture and color information in rash images. Therefore, compared to pixel-difference-based metrics like MAE and MSE, SSIM and HBD demonstrated superior results. Additionally, since our metric evaluates image similarity from multiple dimensions, this means that even if a metric performs poorly or is affected by noise under certain conditions, the other metric might still provide meaningful feedback for the evaluation. 
Therefore, our metric can help MIM achieve optimal performance.
\begin{figure}
    \centering
    \includegraphics[width=\textwidth]{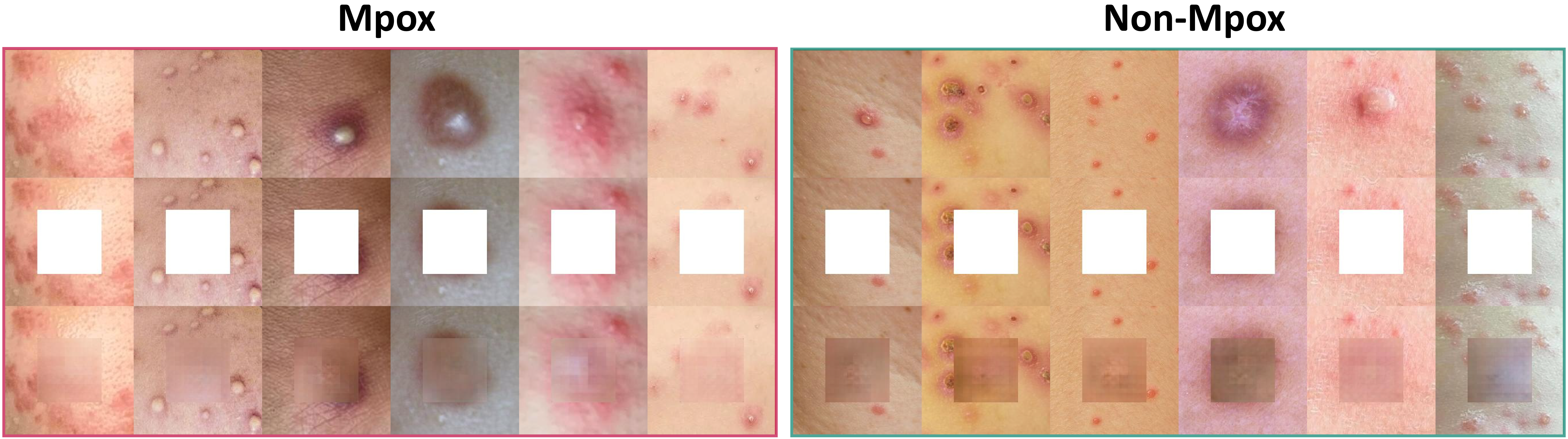}
    \caption{Mpox image inpainted by MIM and non-mpox image inpainted by MIM. The first, second and third columns in the figure represent the original image, the masked image and the repaired image respectively.}
    \label{fig5}
\end{figure}

In order to visually feel the effectiveness of the MIM method, we show some inpainted samples in Figure 5. In Figure 5, we can intuitively see that MIM can better repair the masked mpox image (in-distribution data). Whether in terms of texture, color, or image content, the repair effect of the in-distribution image looks more natural and closer to the original image. On the other hand, for non-mpox images (out-of-distribution data), the repair effect produced by MIM looks very jarring. For example, there is an obvious dissimilar color or a mismatch in image content between the repaired and original images. The visual effect display further proves the rationality of the MIM method.

\subsection{MIM is Better than Classification Models}
One of the motivations for proposing MIM was our belief that the existing classification models were insufficient to detect mpox and non-mpox in real-world settings. Therefore, it is essential to compare MIM with classification models. We selected three representative architectures used for image classification: ResNet (ResNet50)\cite{52}, MobileNetV3 (MobileNetV3 Large)\cite{53}, and Swin Transformer (Swin Transformer Base) \cite{54}.
\begin{figure}
    \centering
    \includegraphics[width=\textwidth]{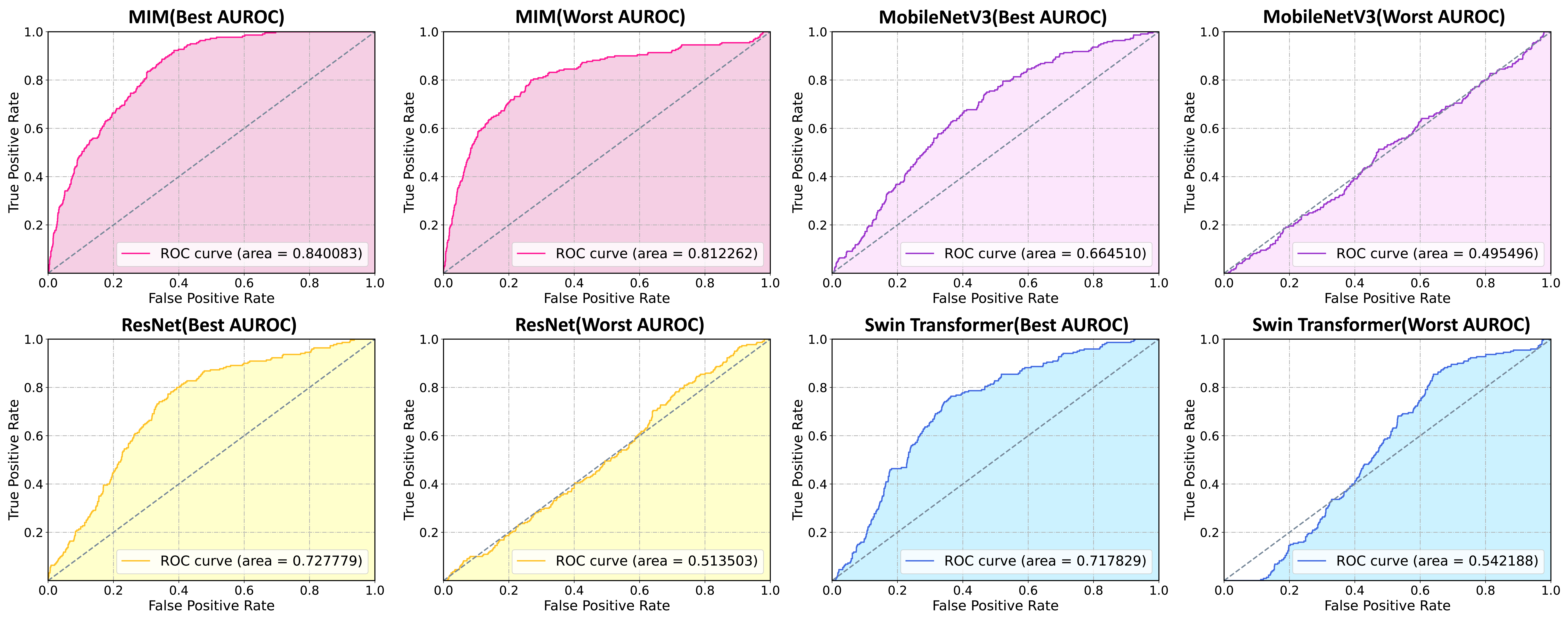}
    \caption{The best AUROC and the worst AUROC of MIM and classification models.}
    \label{fig6}
\end{figure}

Following the experimental setup from related works, we first combined measles, chickenpox, and hand-foot-mouth disease data to create the in-distribution non-mpox data (190 images). Then, we took the remaining fifteen types of skin diseases as the out-of-distribution non-mpox data (1,500 images). Just like with MIM, we used a five-fold cross-validation to train the three classification models above. At this time, the model's in-distribution data will contain four categories. We equally used the average AUROC to assess the potential of classification models in real-world settings. When testing the classification models, in addition to using in-distribution mpox data and non-mpox data, we also tested the models with 1,500 out-of-distribution non-mpox images (anomalous data). Table 3 displays the detailed data usage of classification models. Furthermore, the training parameters for the classification models, such as the number of epochs, optimizer choice, and learning rate, were consistent with what was used for MIM. 
\begin{table}[]
\centering
\caption{Distribution of data used to train and evaluate classification models}
\resizebox{\textwidth}{!}{%
\begin{tabular}{@{}llll@{}}
\toprule
      & Train(ID)                                                                     & Test(ID)                                                                    & Real Test                                                                                                                   \\ \midrule
Fold1 & \begin{tabular}[c]{@{}l@{}}176 mpox images\\ 152 non-mpox images\end{tabular} & \begin{tabular}[c]{@{}l@{}}44 mpox images\\ 38 non-mpox images\end{tabular} & \begin{tabular}[c]{@{}l@{}}(44)→(220) mpox Images (ID)\\ 38 non-mpox images (ID) + 1500 non-mpox Images(OOD)\end{tabular}   \\
Fold2 & \begin{tabular}[c]{@{}l@{}}176 mpox images\\ 152 non-mpox images\end{tabular} & \begin{tabular}[c]{@{}l@{}}44 mpox images\\ 38 non-mpox images\end{tabular} & \begin{tabular}[c]{@{}l@{}}(44)→(220) mpox Images (ID)\\ 38 non-mpox images (ID) + 1500 non-mpox Images(OOD)\end{tabular}   \\
Fold3 & \begin{tabular}[c]{@{}l@{}}176 mpox images\\ 152 non-mpox images\end{tabular} & \begin{tabular}[c]{@{}l@{}}44 mpox images\\ 38 non-mpox images\end{tabular} & \begin{tabular}[c]{@{}l@{}}(44)→(220) mpox Images (ID)\\ 38 non-mpox images (ID) + 1500 non-mpox Images(OOD)\end{tabular}   \\
Fold4 & \begin{tabular}[c]{@{}l@{}}176 mpox images\\ 152 non-mpox images\end{tabular} & \begin{tabular}[c]{@{}l@{}}44 mpox images\\ 38 non-mpox images\end{tabular} & \begin{tabular}[c]{@{}l@{}}(44)→(220) mpox Images (ID)\\ 38 non-mpox images (ID) + 1500 non-mpox Images(OOD)\end{tabular}   \\
Fold5 & \begin{tabular}[c]{@{}l@{}}176 mpox images\\ 152 non-mpox images\end{tabular} & \begin{tabular}[c]{@{}l@{}}44 mpox images\\ 38 non-mpox images\end{tabular} & \begin{tabular}[c]{@{}l@{}}(44)→(220) mpox   Images (ID)\\ 38 non-mpox images (ID) + 1500 non-mpox Images(OOD)\end{tabular} \\ \bottomrule
\end{tabular}
}
\end{table}

Table 4 reports the average AUROC for the three classification models. The Swin Transformer achieved the best results among these three models, with an AUROC of 0.6306. ResNet50 and MobileNetV3 reached 0.6221 and 0.5755, respectively. Since the worst AUROC for the three models is only slightly above 0.5, it can be inferred that, when faced with a large amount of out-of-distribution non-mpox images and augmented mpox images, the classification models are essentially making random predictions. The detection results also showed that these classification models are particularly easy to recognize non-mpox images as mpox. While thoroughly detecting mpox patients can help control the spread of the virus, the model should also recognize its detection performance on non-mpox patients. In real life, the number of non-mpox patients vastly outnumbers mpox patients and a detection method with a high false-positive rate is unacceptable in clinical settings.

\begin{table}[]
\centering
\caption{AUROC of various classification models}
\resizebox{\textwidth}{!}{%
\begin{tabular}{@{}lllllll@{}}
\toprule
Model Name              & Fold1  & Fold2  & Fold3  & Fold4  & Fold5  & Mean(±Std)      \\ \midrule
MobileNetV3   Large     & 0.5745 & 0.5309 & 0.4955 & 0.6645 & 0.6123 & 0.5755(±0.0594) \\
ResNet50                & 0.6342 & 0.6113 & 0.5135 & 0.6239 & 0.7278 & 0.6221(±0.0682) \\
Swin   Transformer Base & 0.7073 & 0.6036 & 0.5422 & 0.5821 & 0.7178 & 0.6306(±0.0698) \\ \bottomrule
\end{tabular}}
\end{table}

Besides, although binary-class classification models do judge some out-of-distribution non-mpox images as non-mpox, this result is difficult for users to understand and convince, as the input to a classification model should belong to in-distribution categories. Compared to classification models, our proposed MIM models the images at the pixel-level. MIM only uses mpox data to train the model, but its average AUROC outperformed Swin Transformer, ResNet, and MobileNetV3 by 0.1931, 0.2016, and 0.2482, respectively, which means MIM is more robust and efficient. Additionally, as MIM only models the mpox images, as MIM only models the mpox images, it can adequately detect non-mpox patients without being influenced by out-of-distribution data. To visually grasp how MIM is more effective than classification models, we plotted the best and worst AUROC of various models in Figure 6.
\subsection{Clinical Testing}
\begin{figure}
    \centering
    \includegraphics[width=\textwidth]{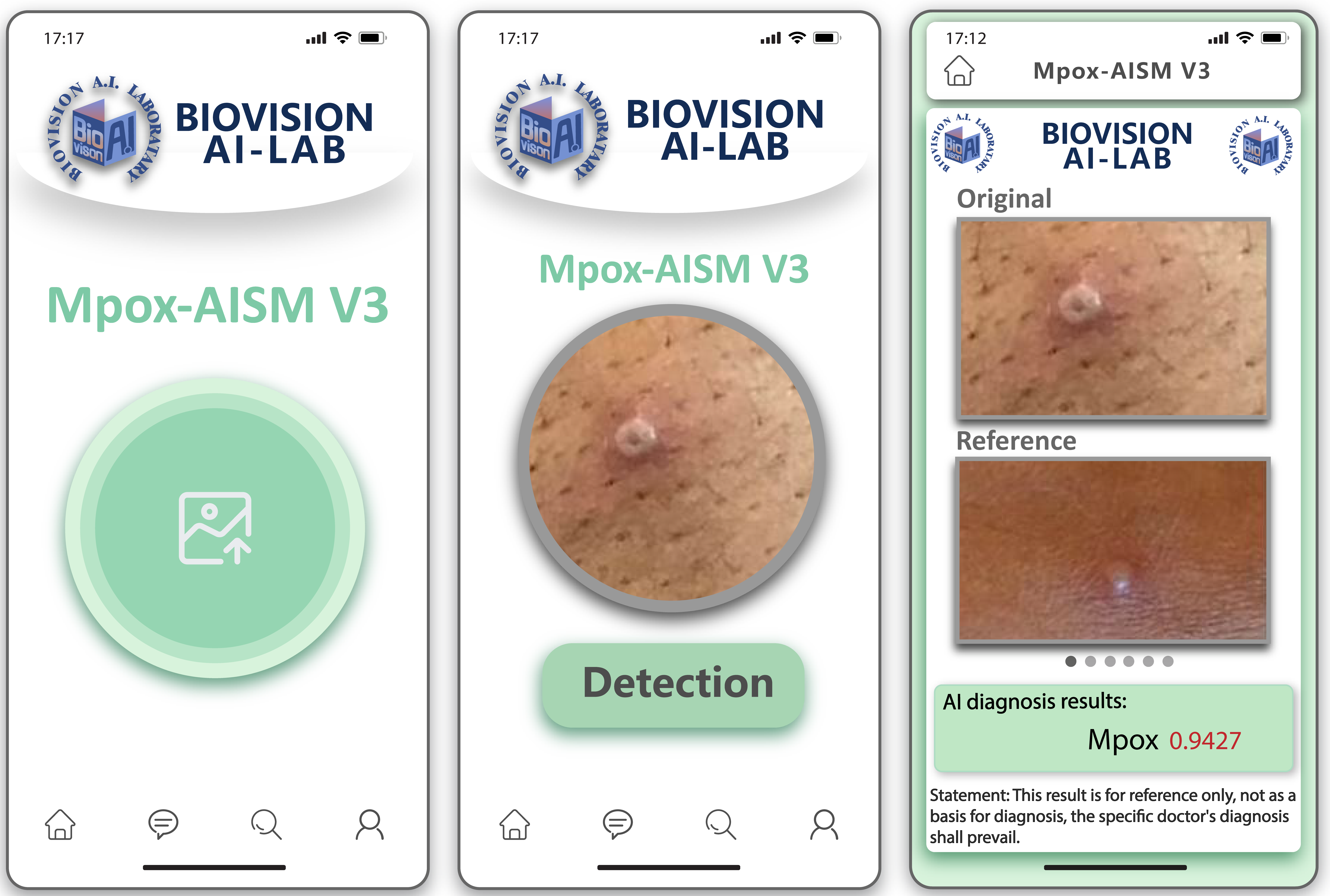}
    \caption{The application page of Mpox-AISM V3.}
    \label{fig7}
\end{figure}
Validating MIM through clinical images provides stronger evidence. We used Gaussian noise to augment the original mpox images in the test set and then obtained 100 new images. This was done to explore the anti-interference ability of MIM and keep the number of mpox images consistent with clinical images. Subsequently, MIM sequentially loaded model weights from fold1 to fold5 and predicted the 100 mpox images and 100 non-mpox images. The results showed that MIM's average AUROC reached 0.881.

Furthermore, these images were also used to test the Swin Transformer, but its average AUROC was only 0.593. We took a closer look at the classification model's prediction results. Consistent with the findings in section 3.3, the Swin Transformer is prone to generating incorrect detection results. This result also confirms that the classification model cannot fully learn the data representation and is, therefore, susceptible to interferences from noise and out-of-distribution data.

The clinical validation results preliminarily validated that the MIM method can effectively identify mpox and non-mpox data in an open-world setting. Simultaneously, it further revealed the shortcomings of classification models. Lastly, based on our previous work, we developed an online smartphone application to provide a more convenient and efficient testing service for people in areas affected by mpox and resource-limited regions \cite{55}. Based on the five-fold cross-validation and clinical validation results, we set the anomaly score threshold to 0.8526 in our application. If MIM calculates a score greater than 0.852 for an image, the app will output "Mpox"; otherwise, it will output "Non-Mpox".

\subsection{Comparison with Related Work}
\begin{table}[]
\centering
\caption{Limitations of related work and strengths of the proposed approach.}
\resizebox{\textwidth}{!}{%
\begin{tabular}{@{}lllllll@{}}
\toprule
Paper    & \begin{tabular}[c]{@{}l@{}}Algorithm\\Name\end{tabular}   & \begin{tabular}[c]{@{}l@{}}Model \\ Type\end{tabular}                   & \begin{tabular}[c]{@{}l@{}}The use of \\ Non-Mpox Data\end{tabular} & Refuse to detect anomaly data  & Robustness in the real-world settings  & \begin{tabular}[c]{@{}l@{}}Clinical \\ Validation\end{tabular} \\ \midrule
{[}13{]} & PoxNet22                                                         & BC classification                                                             & \checkmark                                                                         & ×                                             & Low                                               & ×                                                                    \\
{[}14{]} & \begin{tabular}[c]{@{}l@{}}VGG19\\ MobileNetV2\end{tabular}      & \begin{tabular}[c]{@{}l@{}}BC classification\\ BC classification\end{tabular} & \begin{tabular}[c]{@{}l@{}}\checkmark\\ \checkmark\end{tabular}                             & \begin{tabular}[c]{@{}l@{}}×\\ ×\end{tabular} & \begin{tabular}[c]{@{}l@{}}Low\\ Low\end{tabular} & \begin{tabular}[c]{@{}l@{}}×\\ ×\end{tabular}                        \\
{[}15{]} & DenseNet201                                                      & BC classification                                                             & \checkmark                                                                         & ×                                             & Low                                               & ×                                                                    \\
{[}16{]} & MobileNetV3-s                                                    & BC classification                                                             & \checkmark                                                                         & ×                                             & Low                                               & ×                                                                    \\
{[}17{]} & Monkey-CAD                                                       & BC classification                                                             & \checkmark                                                                         & ×                                             & Low                                               & ×                                                                    \\
{[}18{]} & MobileNetV2                                                      & BC classification                                                             & \checkmark                                                                         & ×                                             & Low                                               & ×                                                                    \\
{[}19{]} & ResNet18                                                         & BC classification                                                             & \checkmark                                                                         & ×                                             & Low                                               & ×                                                                    \\
{[}20{]} & Customed CNN                                                     & BC classification                                                             & \checkmark                                                                         & ×                                             & Low                                               & ×                                                                    \\
{[}21{]} & Vision Transformer                                               & BC classification                                                             & \checkmark                                                                         & ×                                             & Low                                               & ×                                                                    \\
{[}22{]} & \begin{tabular}[c]{@{}l@{}}MobileNet\\ Inception V3\end{tabular} & \begin{tabular}[c]{@{}l@{}}BC classification\\ BC classification\end{tabular} & \begin{tabular}[c]{@{}l@{}}\checkmark\\ \checkmark\end{tabular}                             & \begin{tabular}[c]{@{}l@{}}×\\ ×\end{tabular} & \begin{tabular}[c]{@{}l@{}}Low\\ Low\end{tabular} & \begin{tabular}[c]{@{}l@{}}×\\ ×\end{tabular}                        \\
{[}23{]} & ResNet34                                                         & BC classification                                                             & \checkmark                                                                         & ×                                             & Low                                               & \checkmark                                                                    \\
{[}24{]} & \begin{tabular}[c]{@{}l@{}}Xception\\ ResNet101\end{tabular}     & \begin{tabular}[c]{@{}l@{}}BC classification\\ MC classification\end{tabular} & \begin{tabular}[c]{@{}l@{}}\checkmark\\ \checkmark\end{tabular}                             & \begin{tabular}[c]{@{}l@{}}×\\ ×\end{tabular} & \begin{tabular}[c]{@{}l@{}}Low\\ Low\end{tabular} & \begin{tabular}[c]{@{}l@{}}×\\ ×\end{tabular}                        \\
{[}25{]} & \begin{tabular}[c]{@{}l@{}}Xception\\ DenseNet169\end{tabular}   & \begin{tabular}[c]{@{}l@{}}MC classification\\ MC classification\end{tabular} & \begin{tabular}[c]{@{}l@{}}\checkmark\\ \checkmark\end{tabular}                             & \begin{tabular}[c]{@{}l@{}}×\\ ×\end{tabular} & \begin{tabular}[c]{@{}l@{}}Low\\ Low\end{tabular} & \begin{tabular}[c]{@{}l@{}}×\\ ×\end{tabular}                        \\
{[}26{]} & MonkeyNet                                                        & MC classification                                                             & \checkmark                                                                         & ×                                             & Low                                               & ×                                                                    \\
{[}27{]} & CNN+LSTM                                                         & MC classification                                                             & \checkmark                                                                         & ×                                             & Low                                               & ×                                                                    \\
{[}28{]} & ResNet18                                                         & MC classification                                                             & \checkmark                                                                         & ×                                             & Low                                               & ×                                                                    \\
{[}29{]} & InceptionV3                                                      & MC classification                                                             & \checkmark                                                                         & ×                                             & Low                                               & ×                                                                    \\
{[}30{]} & \begin{tabular}[c]{@{}l@{}}VGG16\\ AlexNet\end{tabular}          & \begin{tabular}[c]{@{}l@{}}MC classification\\ MC classification\end{tabular} & \begin{tabular}[c]{@{}l@{}}\checkmark\\ \checkmark\end{tabular}                             & \begin{tabular}[c]{@{}l@{}}×\\ ×\end{tabular} & \begin{tabular}[c]{@{}l@{}}Low\\ Low\end{tabular} & \begin{tabular}[c]{@{}l@{}}×\\ ×\end{tabular}                        \\
Ours     & MIM                                                              & Generative                                                                    & ×                                                                         & \checkmark                                             & Low                                               & \checkmark                                                                    \\ \bottomrule
\end{tabular}}
\end{table}
Table 5 lists the limitations of previous work and the progress of this work. Since the experimental settings and dataset usage strategies of these works are different, here we focus on discussing the application capabilities of the models in real-world settings. In previous work, most chose a binary-class (BC) classification model to detect monkeypox and non-monkeypox. The remaining work uses multi-class (MC) classification models to detect monkeypox, chickenpox, measles, and normal skin. Although these models have achieved good indicator results, all these classification models require large amounts of non-monkeypox data. Even more fatal is that these classification models always expect the model input to belong to one of the categories in the training set. However, in actual settings, the model more often faces data from out-of-distribution and cannot refuse to detect these data. It is important to reiterate that classification models are easily affected by noise and cannot adequately learn image representations. We have already demonstrated the shortcomings of classification models in the previous section. Compared with these works, our proposed MIM method is more robust and efficient. From a functional point of view, the effect achieved by MIM is consistent with the binary classification model, which is to identify the input as monkeypox or non-monkeypox. Although MIM cannot classify non-monkeypox in more detail, people only care about whether they are infected with monkeypox in daily life. From an algorithmic perspective, MIM only treats monkeypox as normal data and uses generative models such as generative adversarial networks and mask repair tasks to model normal data. MIM cleverly achieves the detection of monkeypox and non-monkeypox without using various non-monkeypox data. Particularly, MIM also avoids the interference of various abnormal data, greatly improving the security and robustness of the model. In order to verify MIM more scientifically, we also performed clinical testing, while most previous work did not perform this verification step. In conclusion, compared with related work, our work has made more rigorous progress through generative models and image inpainting for the first time.

\section{Conclusion}
At the outset of our research, we pointed out the limitations inherent in existing monkeypox detection models. This prompted a pivotal inquiry: Can deep learning effectively differentiate between mpox and non-mpox in real-world settings? To address this, we proposed MIM, drawing inspiration from foundational concepts in One-class Novelty Detection (One-class ND). MIM as a whole is not complicated, but it is effective. It treats monkeypox as normal data and all non-monkeypox data as abnormal, cleverly detecting monkeypox and non-monkeypox. At the same time, MIM obviates the need for non-monkeypox data in classification models and is not affected by abnormal inputs. In addition, MIM also provides a new idea for binary classification tasks in other medical images. For example, MIM may come in handy when detecting whether a lesion exists in a patient's endoscopic or ultrasound image. 

We acknowledge that this work has some limitations. For example, we have yet to explore the impact of different types of generative models on the detection performance of MIM, model complexity, and model response speed. Moreover, the validation and optimization of our model necessitate a more expansive dataset comprising both mpox and non-mpox images. With the recent increase in the number of monkeypox patients, we are strategizing collaborations with specialized medical institutions to procure a more diverse image dataset. We also plan to conduct human-machine comparative studies to benchmark our model's performance against expert evaluations, thereby elevating its reliability and adaptability. Future research focuses on developing lightweight generative detection models. Given its lean architecture, such a model could be seamlessly integrated offline in densely populated areas, such as airports and subways, becoming indispensable tools for large-scale surveillance during outbreaks.
\section*{Data sharing statement}
Data will be made available on request.
\section*{Acknowledgments}
The author is supported by the NSF of Guangdong Province (No.2022A1515011044, No.2023A1515010885), and the project of promoting research capabilities for key constructed disciplines in Guangdong Province (No.2021ZDJS028). Thanks to the Department of Dermatology, the Second Affiliated Hospital of Guangzhou Medical University for their support.
\bibliographystyle{unsrt}  
\bibliography{references}

\end{document}